\documentclass[aps,prstab,superscriptaddress,showpacs]{revtex4}

\usepackage{graphicx}
\usepackage{url}
\usepackage[colorlinks,linkcolor=blue,anchorcolor=blue,citecolor=blue]{hyperref} 
\usepackage{algorithm,algorithmic}
\usepackage{tikz}
\usepackage{suffix}
\usetikzlibrary{arrows,shapes,snakes}
\newcommand{\opal}{\textsc{OPAL}}

\newcommand{\opalcycl}{\textsc{OPAL-cycl}}

\renewcommand{\epsilon}{\varepsilon}

\def\eps{\varepsilon}

\renewcommand {\Re}{{\rm I \kern-2pt R}}

\newcommand {\RM}[1]{\mathrm{#1}}

\newcommand{\bs}[1]{\mathbf #1}

\begin{document}
\tikzstyle{format} = [draw, thin, fill=blue!20]
\tikzstyle{pblock} = [rectangle, draw, fill=blue!20, text width=12em,text centered, rounded corners, minimum height=0.4em]
\tikzstyle{pblockll} = [rectangle, draw, fill=blue!20, text width=25em, text centered, rounded corners, minimum height=0.4em]
\tikzstyle{pblockl} = [rectangle, draw, fill=blue!20, text width=11.5em, text centered, rounded corners, minimum height=0.4em]
\tikzstyle{pblocks} = [rectangle, draw, fill=blue!20, text width=8em, text centered, rounded corners, minimum height=0.4em]
\tikzstyle{decision} = [diamond, draw, fill=blue!20, text width=5em, text badly centered, node distance=3cm, inner sep=0pt]
\tikzstyle{medium} = [ellipse, draw, thin, fill=green!20, minimum height=2.5em]
\tikzstyle{cloud} = [draw, ellipse,fill=red!20, node distance=3cm, minimum height=2em]
\tikzstyle{line} = [draw, -latex']
\tikzstyle{emptyblock} = [rectangle]
\tikzstyle{progblock} = [rectangle, draw, fill=yellow!20, text width=6em, text centered, minimum height=0.4em]
\tikzstyle{null} = [rectangle, fill=blue!0, text width=0em, text centered, rounded corners, minimum height=0.4em]
\tikzstyle{cpoint} = [draw,circle,fill=white,minimum size=1pt,
                            inner sep=0pt]
\tikzstyle{cblock} = [circle, draw, fill=blue!20, text width=3m, text centered, rounded corners, minimum height=0.4em]

\graphicspath{figures}

\title{Towards Quantitative Simulations of High Power Proton Cyclotrons} 

\author{Y. J. Bi}
\email{biyj05@mails.tsinghua.edu.cn}
\affiliation{China Institute of Atomic Energy, Beijing, 102413, China}
\affiliation{Paul Scherrer Institut, Villigen, CH-5232, Switzerland}
\affiliation{Department of Engineering Physics, Tsinghua University, Beijing, 100084, China}
\author{A. Adelmann}
\email{andreas.adelmann@psi.ch}
\author{R. D\"olling}
\author{M. Humbel}
\author{W. Joho}
\author{M. Seidel}
\affiliation{Paul Scherrer Institut, Villigen, CH-5232, Switzerland}
\author{T. J. Zhang}
\affiliation{China Institute of Atomic Energy, Beijing, 102413, China}

\noaffiliation

\begin{abstract}

We describe a large scale simulation effort using \opal~(Object Oriented Parallel Accelerator Library), that leads to a better quantitative understanding of the existing PSI high power proton cyclotron facility.
The 1.3~MW of beam power on target poses stringent constraints on the controlled and uncontrolled beam losses.  We present initial conditions for the Ring simulation, obtained from the new time-structure measurement and the many profile monitors of the 72 MeV transfer line. A trim coil model is developed, needed to avoid the dangerous $\nu_r=2\nu_z$ resonance. By properly selecting the injection position and angle (eccentric injection), the flattop voltage and phase, very good agreement between simulations and measurements at the radial probe RRE4 is obtained. We report on $3 \dots 4$ orders of magnitude in dynamic range when comparing simulations with measurements. The relation between beam intensity, rms beam size, and accelerating voltage is studied and compared with measurement. The demonstrated capabilities are mandatory in the design and operation of the next generation high power proton drivers. In an outlook we discuss our future plans to include more physics into the model, which eventually leads to an even larger dynamic range
in the simulation.

\end{abstract}


\maketitle

\section{INTRODUCTION \label{intro}}
PSI operates a cyclotron based high intensity proton accelerator routinely at an average beam power of 1.3 MW. With this power the facility is at the worldwide forefront of high intensity proton accelerators. 
An upgrade program is under way to ensure high operational reliability and push the intensity to even higher levels. The beam current is limited in practice by losses at extraction and the resulting activation of accelerator components. 
Further intensity upgrades and new projects aiming at an even higher average beam power, are only possible if the relative losses can be lowered in proportion, thus keeping absolute losses at a constant level. 

Maintaining beam losses at levels allowing hands-on maintenance is a primary challenge in any high power proton machine design and operation. For a $1.3$ MW beam in the PSI Ring cyclotron this corresponds to a transmission of $99.97\%$ taking controlled and uncontrolled losses into account. 
In a 10 MW class machine we require the losses to be on the same level which is a challenging task and is asking for precise beam dynamics calculation.  In consequence,  predicting beam halo at these levels is a great challenge and will be addressed in this paper. 

High power hadron drivers have being used in many disciplines of science and, a growing interest in cyclotron technology for high power hadron drivers has be shown very recently. Two very recent papers demonstrate this fact: 
1) The search for $CP$ violation in the Neutrino sector \cite{PhysRevLett.104.141802}
calls ultimately for three machines in the megawatt range at an energy of 800 MeV. 2) in \cite{adswhitepaper2010}, a white paper on {\it Accelerator and Target Technology for Accelerator Driven Transmutation and Energy Production} the cyclotron technology 
is advertised, quote: {\em On the whole, the development status of accelerators is well advanced, and beam powers 
of up to 10 MW for cyclotrons and 100 MW for linacs now appear to be feasible ....}.

This report will briefly introduce \opal, a tool for precise beam dynamics simulations including 3D space charge. One of \opal's "flavors" (\opal-cycl) is dedicated to high power cyclotron modeling and is explained in greater detail. We then explain how to obtain
initial conditions for our PSI Ring cyclotron which still delivers the world record in beam power of 1.3 MW in continuous wave (cw) operation. Several crucial steps are explained, necessary to be able to predict tails at the level of $3\sigma \dots 4\sigma$ in the PSI Ring cyclotron. We compare
our results at the extraction with measurements, obtained with a $1.18$ MW cw production beam. Based on measurement data, we develop a simple linear model to predict beam sizes of the extracted beam as a function of intensity and confirm the model with simulations.
A conclusion and discussions to include more physics into the model, which eventually leads to a even larger dynamic range in the simulation, closes the paper.

\section{BASIC EQUATIONS AND PHYSICAL MODEL }

\subsection{A BRIEF LOOK AT \opal}

\opal~(Object Oriented Parallel Accelerator Library) is a tool for charged-particle optic calculations in large accelerator structures and beam lines including 3D space charge. \opal~ is built from first principles as a parallel application, it admits simulations of any scale: on the laptop and up to the largest High Performance Computing (HPC) clusters available today. Simulations, in particular HPC simulations, form the third pillar of science, complementing theory and experiment. \opal~ includes various beam line element descriptions and methods for single particle optics, namely maps up to arbitrary order, symplectic integration schemes and lastly time integration \cite{opal:1}. \opal~ is based on IPPL (Independent Parallel Particle Layer) \cite{ippl:1} which adds parallel capabilities. Main functions inherited from IPPL are: structured rectangular grids, fields and parallel FFT and particles with the respective interpolation operators. Recently we added a powerful iterative solver to \opal\ taking into account complicated boundary conditions \cite{Adelmann20104554}.  More details on cyclotron modeling which are direct relevant to this article can be found in \cite{PhysRevSTAB.13.064201}. Several flavors of \opal\ are
available. For details we refer to the User Manual  \cite{opal:1}. In this paper we use \opal-t\ for the tracking of 72 MeV beam line, connecting two cyclotrons, the Injector 2 and the
Ring Cyclotron. The other \opal\ flavor -  \opal-cycl\ - is designed specially for cyclotron beam dynamics and, is explained in the next section. 

\subsection{THE BEAM DYNAMICS MODEL OF \opal-cycl}
In the cyclotrons and beam lines under consideration, the collisions between beam particles can be neglected because the typical bunch densities are low.
In time domain, the general equations of motion of a charged particle in electromagnetic fields can be expressed by
\begin{equation}\label{eq:motion}
  \frac{d\bs{p}(t)}{dt}  = q\left(c\mbox{\boldmath$\beta$}\times \bs{B} + \bs{E}\right), \nonumber \\
\end{equation}
where $m_0, q,\gamma$ are rest mass, charge and the relativistic factor. With $\bs{p}=m_0 c \gamma \mbox{\boldmath$\beta$}$ we denote the momentum of a particle,
$c$ is the speed of light, and $\mbox{\boldmath$\beta$}=(\beta_x, \beta_y, \beta_z)$ is the normalized velocity vector. In general the time ($t$) and position ($\bs{x}$) dependent electric and magnetic vector fields are
written in abbreviated form as $\bs{B} \text{ and } \bs{E}$.

If $\bs{p}$ is normalized by $m_0c$,
Eq.\,(\ref{eq:motion}) can be written in Cartesian coordinates as
\begin{eqnarray}\label{eq:motion2}
  \frac{dp_x}{dt} & = & \frac{q}{m_0c}E_x + \frac{q}{\gamma m_0}(p_y B_z - p_z B_y),    \nonumber \\
  \frac{dp_y}{dt} & = & \frac{q}{m_0c}E_y + \frac{q}{\gamma m_0}(p_z B_x - p_x B_z),   \\
  \frac{dp_z}{dt} & = & \frac{q}{m_0c}E_z + \frac{q}{\gamma m_0}(p_x B_y - p_y B_x).    \nonumber
\end{eqnarray}
The evolution of the beam's distribution function $ f(\bs {x},c\mbox{\boldmath$\beta$},t)$ can be expressed by a collisionless Vlasov equation:
\begin{equation}\label{eq:Vlasov}
  \frac{df}{dt}=\partial_t f + c\mbox{\boldmath$\beta$} \cdot \nabla_x f +q(\bs{E}+ c\mbox{\boldmath$\beta$}\times\bs{B})\cdot \nabla_{c\mbox{\boldmath$\beta$}} f  =  0,
\end{equation}
where $\bs{E}$ and $\bs{B}$ include both external applied fields, and space charge fields
\begin{eqnarray}\label{eq:Allfield}
  \bs{E} & = & \bs{E_{\RM{ext}}}+\bs{E_{\RM{sc}}}, \nonumber\\
  \bs{B} & = & \bs{B_{\RM{ext}}}+\bs{B_{\RM{sc}}}.
\end{eqnarray}
In order to model a cyclotron, the external electromagnetic fields are given by measurements or by numerical calculations.

The space charge fields can be obtained
by a quasi-static approximation. In this approach, the relative motion of the particles is non-relativistic in the beam rest frame, so the self-induced magnetic field is practically absent and the electric field can be computed by solving Poisson's equation
\begin{equation}\label{eq:Poisson}
  \nabla^{2} \phi(\bs{x}) = - \frac{\rho(\bs{x})}{\varepsilon_0},
\end{equation}
where $\phi$ and $\rho$ are the electrostatic potential and the spatial charge density in the beam rest frame. The electric field can then be calculated by
\begin{equation}\label{eq:Efield}
  \bs{E}=-\nabla\phi,
\end{equation}
and back transformed to yield both the electric and the magnetic fields, in the lab frame, required in Eq.\,(\ref{eq:Allfield}) by means of a Lorentz transformation.
Because of the large vertical gap in our cyclotron, the contribution of image charges and currents are minor effects compared to space charges \cite{Baartman:1}, and hence it is a good approximation to use
open boundary conditions.

The combination of Eq.\,(\ref{eq:Vlasov}) and Eq.\,(\ref{eq:Poisson}) constitutes the Vlasov-Poisson system.
In the following, the method of how to solve these equations in cyclotrons using PIC methods is described in detail.

\begin{figure}
    {\includegraphics[width=8cm]{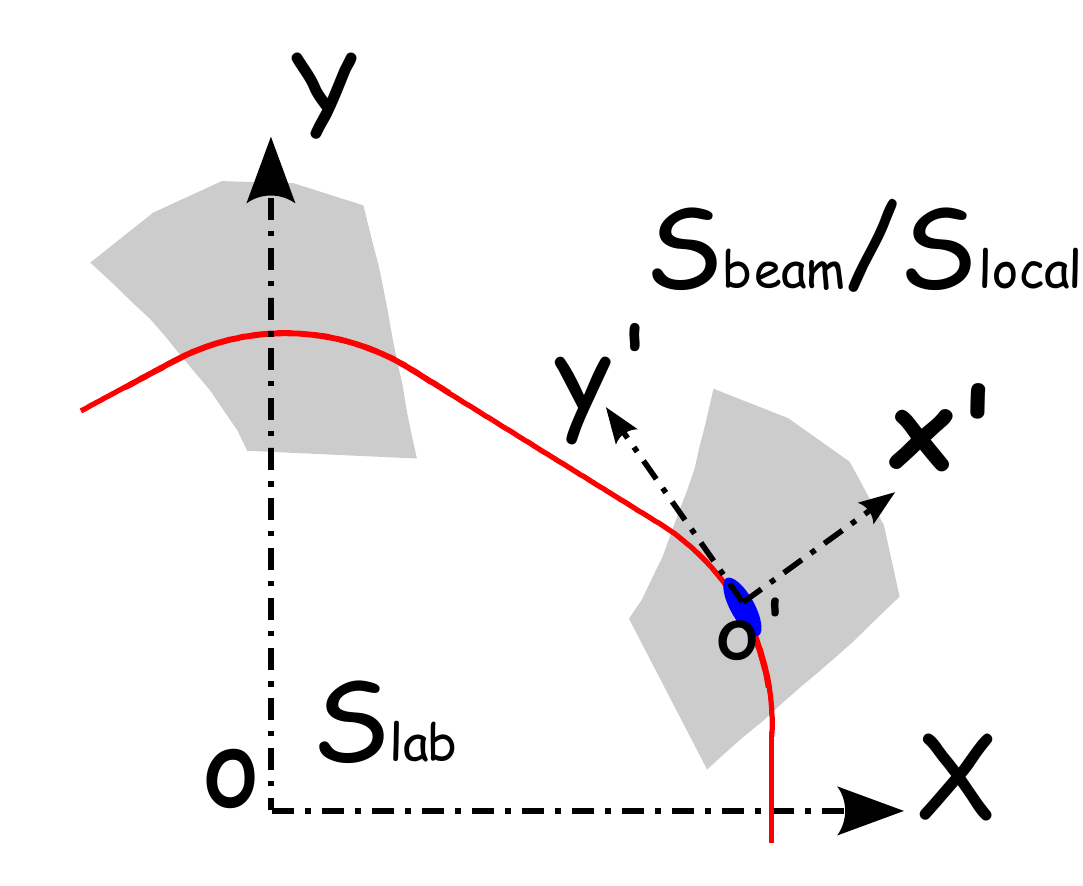}}
    \caption{(Color) Schematic plot of the top view of the three coordinate frames. The red curve is the orbit of the bunch center, 
      the blue area represents the bunch shape, and the gray area is the hill region of magnetic field.}
    \label{fig:frame}
  \end{figure}

Considering that particles propagate spirally outwards in cyclotrons, and the longitudinal orientation changes continuously,
three right-handed Cartesian coordinate systems are defined, as shown in Fig.\,\ref{fig:frame}.
The first coordinate system is the fixed laboratory frame ${\bs{S}_{\RM{lab}}}$, in which the external field data is stored and the particles are tracked.

Its origin is the center of the cyclotron and its $X-Y$ plane is the median plane and the positive direction of $Z$ axis points vertically upwards.

The second coordinate system is the local instantaneous frame ${\bs{S}_{\RM{local}}}$, which is a temporal auxiliary frame for the space charge solver.
Its origin $O'$ is the mass center of the beam and the orientation of the $Y'$ axis is coincident with the average longitudinal direction and
the positive orientation of the $Z'$ axis points vertically upwards.

The third coordinate system is the beam rest frame $\bs{S}_{\RM{beam}}$, which is co-moving with the centroid of the beam.
It has the same orientation and origin as ${\bs{S}_{\RM{local}}}$, but the length in longitudinal
direction is scaled by $1/\gamma$ due to relativistic effects.

At each time step, the frames $\bs{S}_{\RM{local}}$ and $\bs{S}_{\RM{beam}}$ are redefined according to the current 6D 
phase space distribution, and all particles are transformed from $\bs{S}_{\RM{lab}}$ to $\bs{S}_{\RM{local}}$, 
then a Lorentz transformation is performed to transform all particles to $\bs{S}_{\RM{beam}}$.
The Poisson equation is then solved in the frame $\bs{S}_{\RM{beam}}$. In a 3D Cartesian frame, the solution of the Poisson equation at point $(x,y,z)$ can be expressed by
\begin{equation}\label{eq:Poten}
  \phi(x,y,z)= \frac{1}{4\pi\varepsilon_0}\int{G(x,x',y,y',z,z')\rho(x',y',z')dx'dy'dz'},
\end{equation}
with $G$ the 3D Green function
\begin{equation}\label{eq:Green}
  G(x,x',y,y',z,z')= \frac{1}{\sqrt{(x-x')^2+(y-y')^2+(z-z')^2}},
\end{equation}
assuming open boundary conditions. Details of the space charge field calculation can be found in \cite{Hockney:1}.

The model of the external magnetic field is based on mid-plane field measurements with excited trim coils. In consequence we have a vertical field, $B_z$,  measured on the median plane ($z=0$) as a
function of azimuthal position ($\theta$).
Since the magnetic field outside the median plane is required to compute trajectories with $z \neq 0$, the field needs to be expanded in the $Z$ direction.
According to the approach given by Gordon and Taivassalo \cite{Gordon:2}, by using a magnetic potential and the measured $B_z$ on the median plane
at the point $(r,\theta, z)$ in cylindrical polar coordinates, the 3$rd$ order field can be written as
\begin{eqnarray}\label{eq:Bfield}
  B_r(r,\theta, z) & = & z\frac{\partial B_z}{\partial r}-\frac{1}{6}z^3 C_r, \nonumber\\
  B_\theta(r,\theta, z) & = & \frac{z}{r}\frac{\partial B_z}{\partial \theta}-\frac{1}{6}\frac{z^3}{r} C_{\theta}, \\
  B_z(r,\theta, z) & = & B_z-\frac{1}{2}z^2 C_z,  \nonumber
\end{eqnarray}
where $B_z\equiv B_z(r, \theta, 0)$ and
\begin{eqnarray}\label{eq:Bcoeff}
  C_r & = & \frac{\partial^3B_z}{\partial r^3} + \frac{1}{r}\frac{\partial^2 B_z}{\partial r^2} - \frac{1}{r^2}\frac{\partial B_z}{\partial r}
        + \frac{1}{r^2}\frac{\partial^3 B_z}{\partial r \partial \theta^2} - 2\frac{1}{r^3}\frac{\partial^2 B_z}{\partial \theta^2}, \nonumber  \\
  C_{\theta} & = & \frac{1}{r}\frac{\partial^2 B_z}{\partial r \partial \theta} + \frac{\partial^3 B_z}{\partial r^2 \partial \theta}
        + \frac{1}{r^2}\frac{\partial^3 B_z}{\partial \theta^3},  \\
  C_z & = & \frac{1}{r}\frac{\partial B_z}{\partial r} + \frac{\partial^2 B_z}{\partial r^2} + \frac{1}{r^2}\frac{\partial^2 B_z}{\partial \theta^2}. \nonumber
\end{eqnarray}

All the partial differential coefficients are computed on the median plane data by interpolation, using Lagrange's 5-point formula.

Finally both the external fields and space charge fields are used to track particles for one time step using a 4$th$ order Runge-Kutta (RK) integrator, in which
the fields are evaluated for four times in each time step. Space charge fields are assumed to be constant during one time step,
because their variation is typically much slower than that of external fields. More details and unique features can be found in \cite{PhysRevSTAB.13.064201}.

\section{OBTAINING INITIAL CONDITIONS FOR THE RING CYCLOTRON}
\begin{figure}
    {\includegraphics[width=0.95\linewidth]{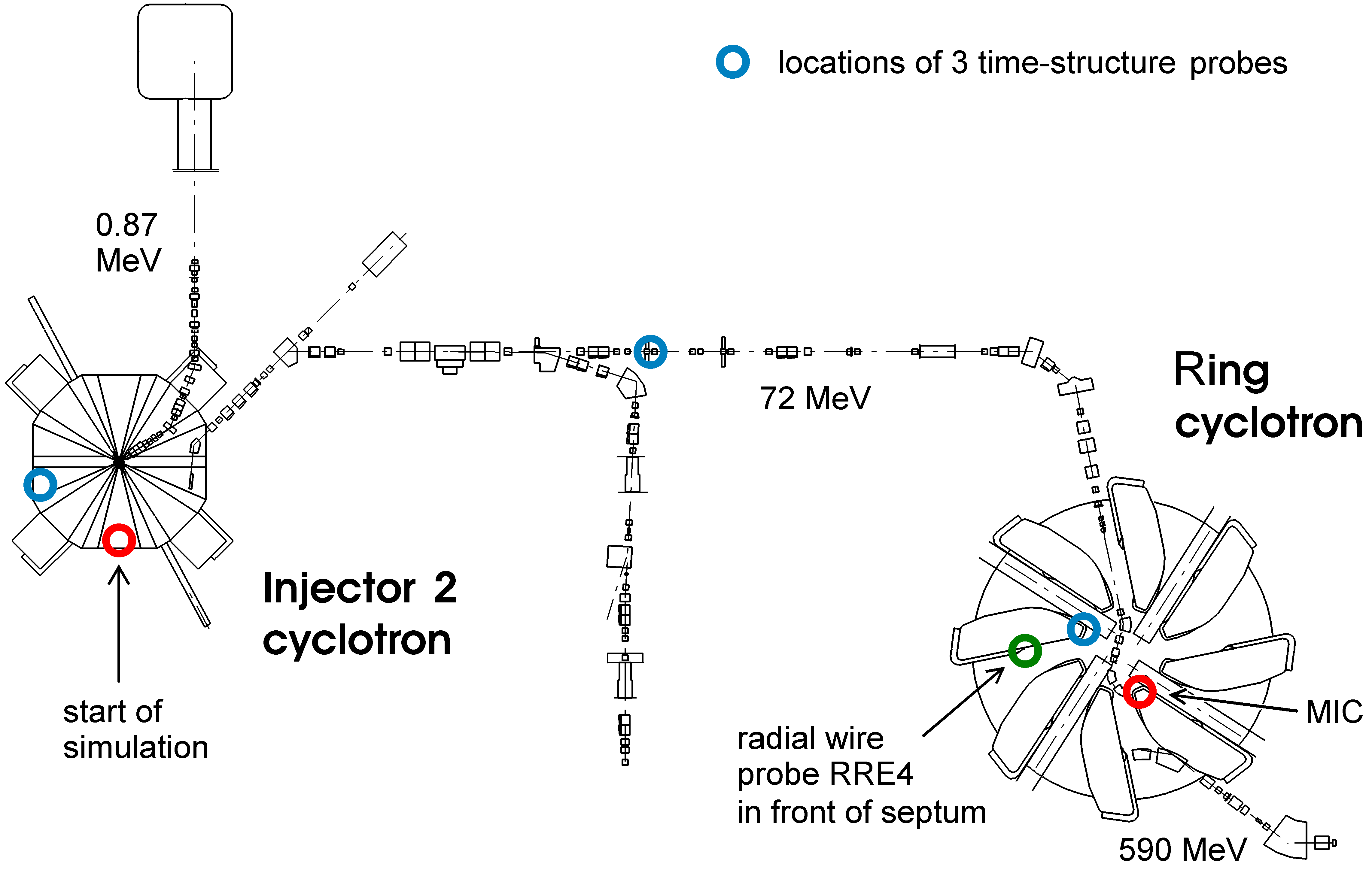}}
    \caption{(Color) The PSI Injector 2 cyclotron and beam transfer line to the PSI Ring cyclotron.}
    \label{fig:overview}
  \end{figure}

At the extraction region of the Injector 2 we only have a very limited number of measurement data, however in the injecting beam line connecting the two cyclotrons we 
have $14$ vertical and $17$ horizontal beam profile monitors available for high intensity operation. Three time-structure measurements, one at the last turn of the Injector 2, one 27 meters 
downstream and one at the first turn of the PSI Ring cyclotron give important information on the longitudinal beam size \cite{timestructure}.  
In an overview (Fig \ref{fig:overview}) the starting point of the simulations, and some of the diagnostics are shown. We note that from a beam dynamics point of view, the particles travel in the order of $4$ km, from the 
marked start of the simulation to the RRE4, the probe covering the last 9 turns of the PSI Ring cyclotron.

We start the \opal-t\ simulations from the middle of the last valley before extraction from the Injector 2 and perform a full 3D simulation until the magnetic injection channel (MIC)
of the PSI Ring cyclotron. In Tab. \ref{tab:injline} the initial values for the simulation of important beam parameters are shown. At the MIC we resample the distribution and switch to \opal-cycl\ for the PSI Ring cyclotron simulation. 
The new distribution is sampled using the moments obtained from the transfer line simulation (at MIC).

\begin{table}[h]\footnotesize
{\renewcommand{\arraystretch}{1.5}
\renewcommand{\tabcolsep}{0.5cm}}
\caption{Initial conditions of the 72 MeV transfer line for a 2 mA cw beam. The emittances are non-normalized. }
\centering
  \label{tab:injline}
  \begin{tabular}{  l l l l l l l l l l }
    Distribution & $\epsilon_x$  (mm-mrad) & $\epsilon_y$  (mm-mrad)  & $x_{rms} $ (mm)  &  $y_{rms} $ (mm) &  $l_{rms} $ (mm) & $\delta$ (\%) &  $\langle x x' \rangle$ &  $\langle y y' \rangle$&  $\langle x \delta\rangle$ \\
     \hline
    3D Gaussian & 2.22 & 0.43  & 2.9 & 0.5 &6.2&0.06& -0.14 & 0.07 &-0.92\\
      \hline 

  \end{tabular}
 \end{table}

\begin{figure}[H]
   \centering
  \includegraphics*[width=0.95\linewidth]{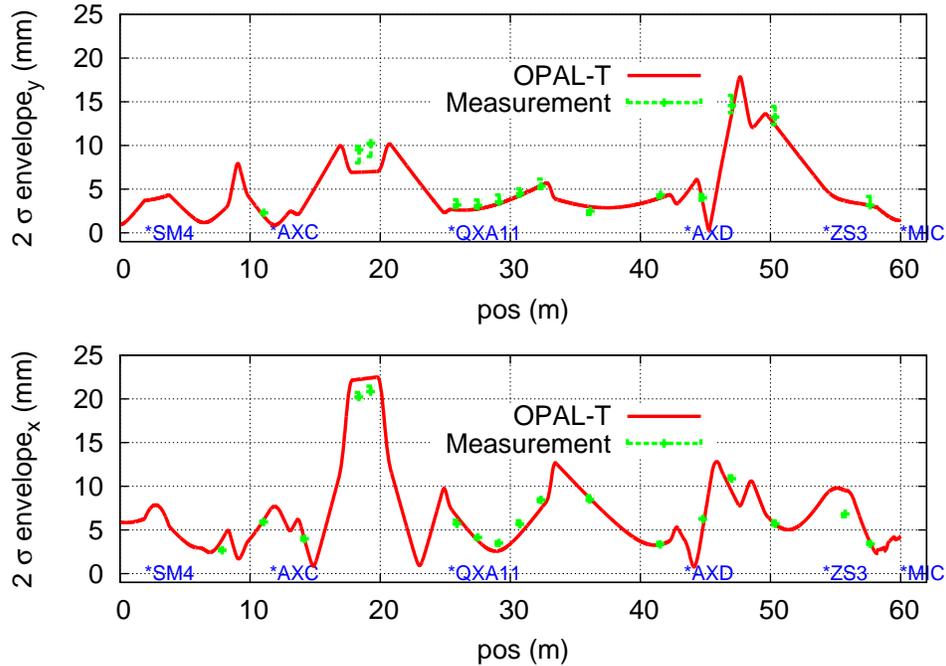}
   \caption{Envelope of the beam in the 72 MeV transfer line for  a 2 mA beam. The error bars showing different measurements at the same intensity level}
   \label{fig:envelope}
\end{figure}

\begin{figure}[H]
   \centering
  \includegraphics*[width=0.95\linewidth,angle=0]{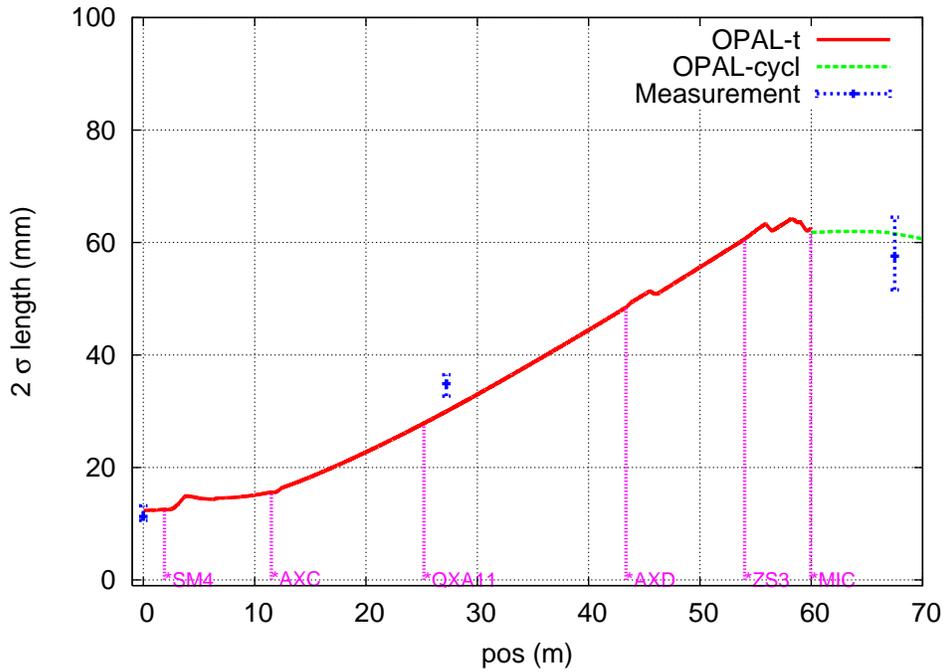}
   \caption{Bunch length of the beam in the 72 MeV transfer line for a 2 mA beam (in Fig \ref{fig:overview} the positions of the time-structure measurements are shown).}
   \label{fig:length}
\end{figure}

Figure \ref{fig:envelope} shows the comparison of the  $2\sigma$ beam width between \opal-t\ and the measurements.  The model predicts very well the evolution of the envelope from the beginning to the end of the transfer line.
Figure \ref{fig:length} shows the comparison of the predicted bunch length by the model and the measurements using the time-structure probes. The large error bar at the 

injection to the Ring is because of the large background.  The longitudinal initial conditions, in the center of the valley between sector magnet 3 and 4 (see Fig: \ref{fig:overview}), are derived from the time structure measurement, which is $\sim 4$~m upstream. 
In the center of the valley, the major axis of the bunch ellipse is along the longitudinal direction, and hence we obtain the transformed initial condition by a simple rotation in the longitudinal and radial plane.

%
\section{TOWARDS REALISTIC HIGH POWER CYCLOTRON SIMULATIONS}

The beam losses during the operation of the cyclotron usually limits the intensity that can be extracted.
The PSI 590 MeV Ring routinely delivers 2.2 mA of cw beam, having a very low integrated loss rate, of the order of 0.02\%. This thight margin avoids excessive activation of accelerator components and hence keeps the radiation dose imposed on the personnel involved in maintenance at acceptable levels. Furthermore, about  $90$\% of the losses are located at injection and extraction.

Therefore, the understanding of the beam dynamics and, the knowledge of the detailed beam distribution especially at the extraction region, is one of the key points to be addressed especially if power levels increase
in future projects \cite{adswhitepaper2010}.

Several important effects which need to be carefully modeled to keep extraction losses in the order of 0.02\% are:

\begin {itemize}
\item the turn separation at the position of the extraction septum must be made as large as possible,
\item the radial beam size at the extraction region must be smaller than the turn separation,
\item the halo, especially at the extraction, has to be minimized,
\item in case of the PSI Ring cyclotron,  a long "pencil" beam is used and hence the linear space charge effects must be effectively compensated to avoid the formation of a S-shaped beam which apparently increases the effective radial beam size.
\end {itemize}
We now discuss these issues related to the PSI Ring cyclotron which however can be considered universal for high power cyclotrons, and hence are certainly important for future high intensity related projects
 \cite{adswhitepaper2010}.
\subsection{The Flattop Phase}

Although a compact beam is observed at the extraction of the Injector 2 cyclotron, the bunch length increases from about $\sigma=6$ $mm$ to about $\sigma=31$ $mm$ 
at injection into the Ring after passing through the almost 60 m long (72 MeV) transfer line.
For such a long "pencil" beam, a flattop cavity is needed to compensate the energy difference from the main cavity and avoid the formation of the S-shape beam caused by space charge effects.

When there is no space charge effect, the ideal flattop makes the total energy gain
of any particle almost the same independent of the RF phase.
Considering a high current beam, the flattop phase must be shifted such that the tail particles gain more energy than the head particles. This compensates exactly the linear part of the space charge force. Therefore the phase of the flattop is adjusted intensity-dependent and, there exists an optimum flattop phase for a given intensity.

For our simulation we use $11.5$\% of the sum of the main cavity voltages as the flat-top cavity voltage (as set in the control room)  and adjust the phase to obtain the same phase difference between main and flat-top cavity as set in the control room.

\subsection{The Effect of the Trim Coil TC15}

\begin{figure}[H]
   \centering
  \includegraphics*[width=0.95\linewidth]{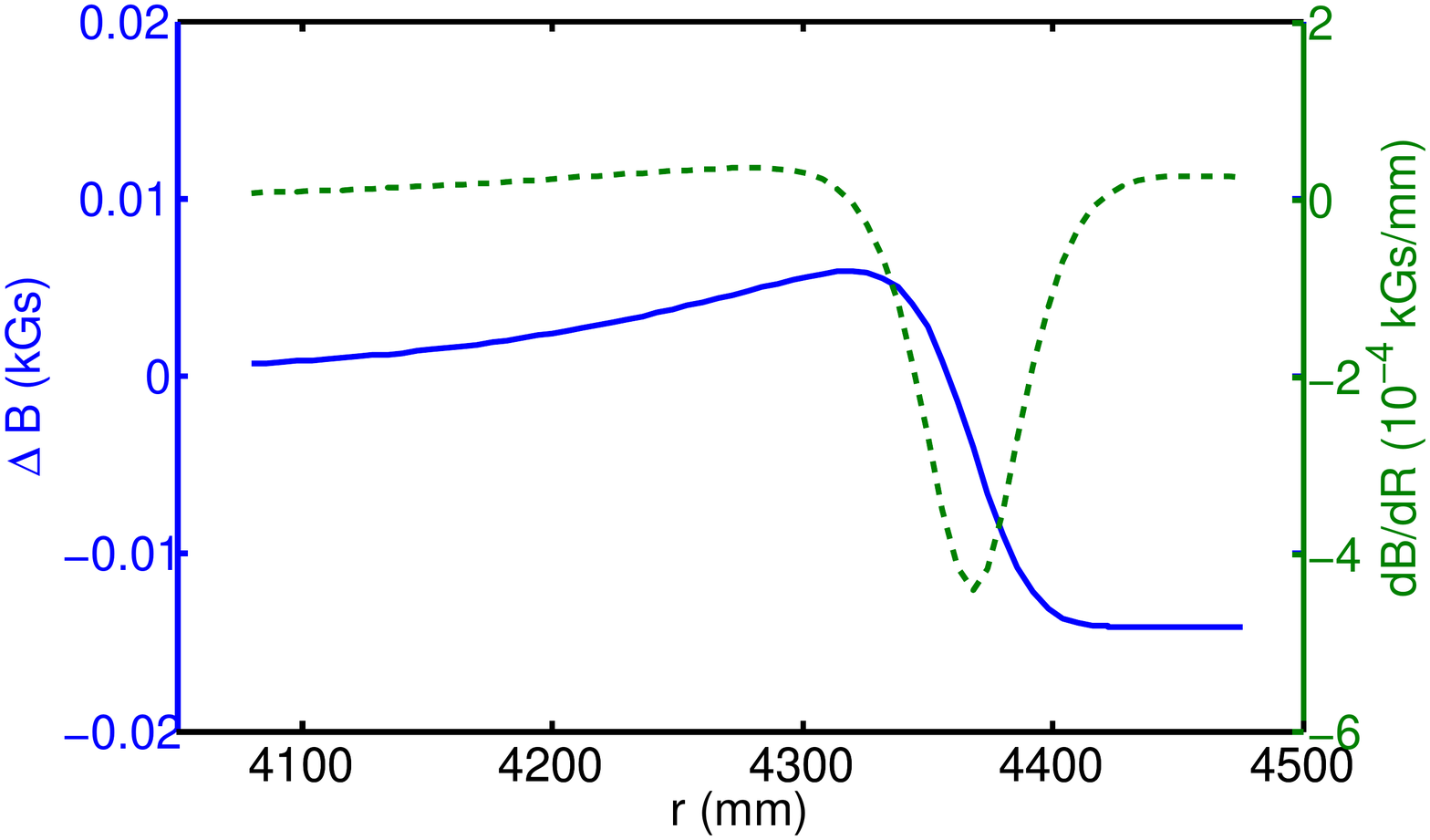}
   \caption{The field (solid line) and field gradient (dashed line) of TC15 in the PSI Ring cyclotron.}
   \label{fig:trimcoil}
\end{figure}

A small manufacturing error produces a slight deviation in the average field profile and a corresponding shift in the tunes $\nu_r$ and $\nu_z$. This requires a strong excitation of the 
trim coil TC15. Without this correction, the coupling resonance $\nu_r=2\nu_z$ would be crossed four times at energies of 490, 525, 535 and 585 MeV, respectively.  A large horizontal oscillation would be transformed into a large vertical one at the coupling resonance which can lead to large vertical beam losses.
An analytic model was developed which mimics the field due to real trim coil characteristics  \cite{TRIMCOIL}.
It is described by Eq.\,(\ref{eq:tcmodel}).
\begin{equation}\label{eq:tcmodel}
\Delta B=-B_a \left[ A_1+\frac{A_2}{10^{a_1 R+b_1}+1}+\frac{A_3}{10^{a_2 R+b_2}+1}\right],
\end{equation}
where R is the radius, $B_a$ is the maximum magnetic field, 
the constants $A_1=-1.08$, $A_2=1.08$, $A_3=1.80$, $a_1=0.005$,
$b_1=-21.72$, $a_2=-0.033$, $b_2=145.60$ for our case.
It provides an additional magnet field and field gradient in the radial direction as shown in Fig. \ref{fig:trimcoil}.

\begin{figure}[H]
   \centering
  \includegraphics*[width=0.95\linewidth]{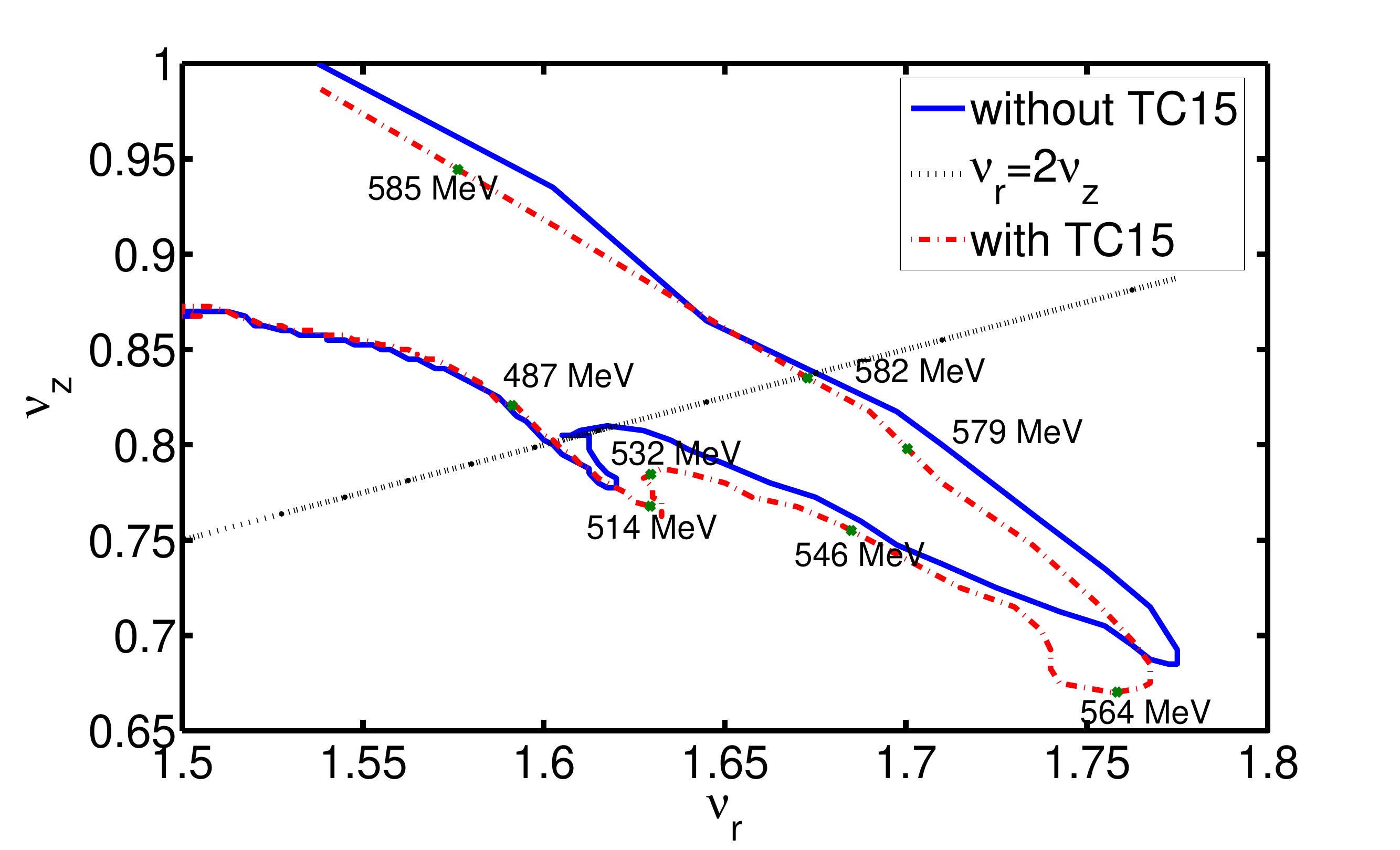}
   \caption{Tune diagram with and without TC15.}
   \label{fig:nurnuzpart}
\end{figure}

The trim coil gives a maximum magnetic field of 14 Gauss, and furthermore has a long tail towards smaller radii in order to make the integrated strength of the trim coil over the radius to zero.
The radial and vertical tune shifts caused by TC15 are given by,
\begin{equation}
\left\{
\begin{array}{c}
 \Delta\nu_r\approx \frac{R}{2\nu_r B}\frac{d\overline{B}}{dR}\approx0.014\\
 \Delta \nu_z\approx -\frac{\nu_r}{\nu_z}\Delta\nu_r\approx-2\Delta\nu_r\approx-0.028
\end{array}
 \right.
 \end{equation}
where $R$ is the orbit radius, $B$ is the hill field, $\frac{d\overline{B}}{dR}$ is the average field gradient in radial direction. Careful beam dynamics studies have shown the meaningfulness  of such detailed modeling in order to obtain a complete and precise pictures of the beam dynamics in the PSI Ring cyclotron. The modified tune diagram by TC15 is shown in figure. \ref{fig:nurnuzpart}.
Without TC15, in simulations and in the operation of the Ring, we observe severe vertical beam losses and can not obtain the required extraction efficiency.

\subsection{The Injection Position and Angle}
\begin{figure}[H]
   \centering
  \includegraphics*[width=0.95\linewidth]{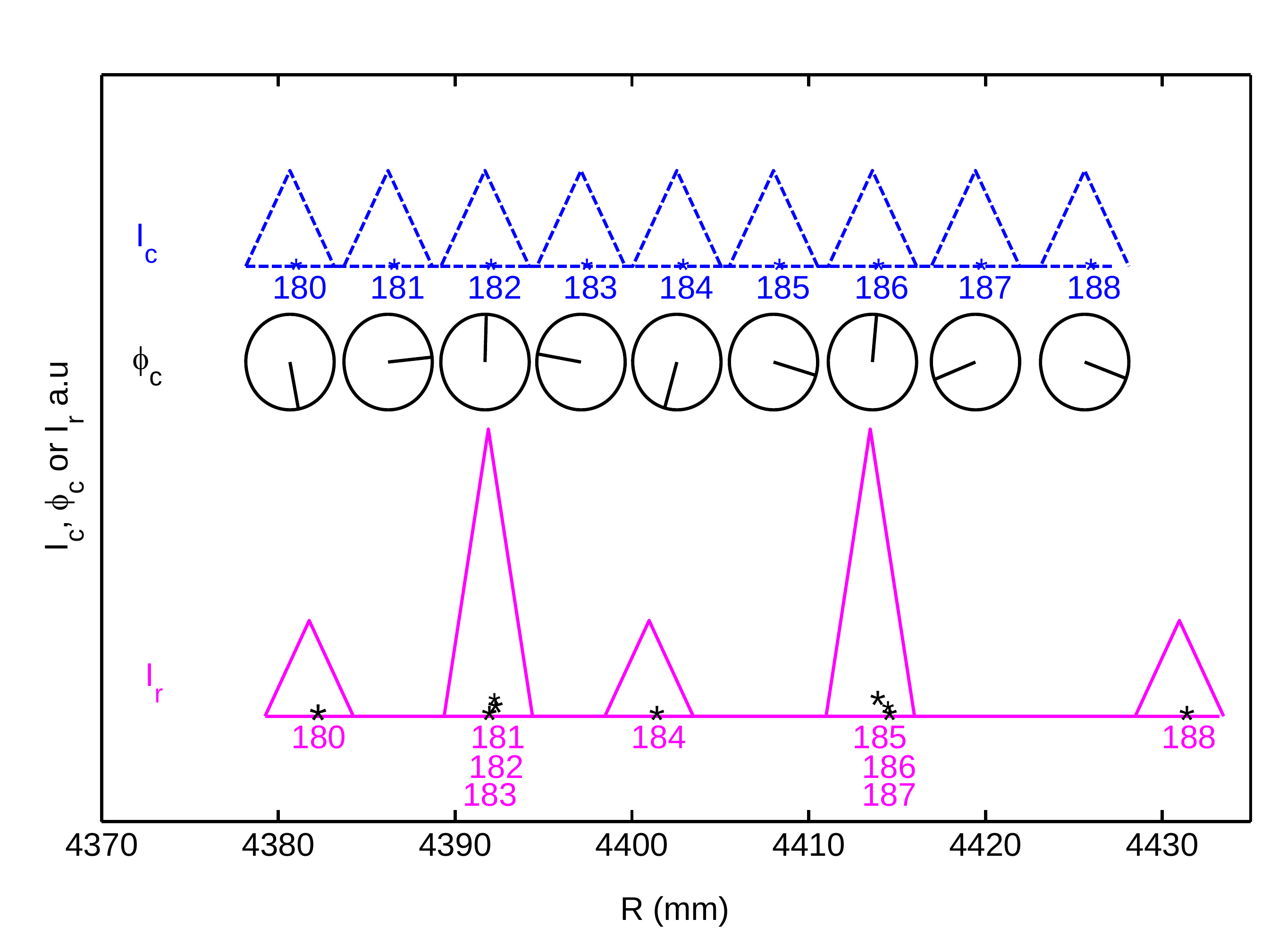}
   \caption{Schematic representation of the turn pattern in the PSI Ring cyclotron. $I_c$, $\phi_c$ and $I_r$ represent the intensity distribution of a centered beam with $6$ mm turn separation, the betatron oscillation phase of an eccentric beam and the intensity distribution of a real beam with eccentric injection.}
   \label{fig:turnpattern}
\end{figure}

The PSI Ring cyclotron has a single turn extraction, hence a large radial turn separation between the last two turns is required.  
The turn separation for a centered beam is defined as
\begin{equation}
\label{eq:dRdn} \frac{dR}{dn}=\frac{\gamma}{\gamma+1}R\frac{dE/dn}{E}\frac{1}{1+k},
\end{equation}
where $k$ is the field index. For the PSI Ring cyclotron this gives about $6.0$ mm (Fig. \ref{fig:turnpattern} upper part)  at the extraction region, which is not enough for 
high current operation and would result in large losses.

To increase the turn separation, a non-centered injection into the PSI Ring cyclotron is used. Since $\nu_r\approx 1.7$ at extraction,
adjusting the injection position and angle, results in the betatron amplitude being almost equal to the increase in radius per turn.
The formation of the turn pattern under this condition, for the last nine turns, is shown in Fig. \ref{fig:turnpattern} (lower part).
This is a special turn pattern because the last turn is well separated from the overlapping second, third and fourth last turns. In this case, the turn separation at the extraction turn is as large as $16$ mm, hence it allows the extraction of a high intensity beam with very low losses.

\subsection{Comparing the Radial Intensity Profile at Extraction with Measurements}
Up to now we have described the most important  steps in setting up a precise beam dynamics simulation of the PSI Ring cyclotron. We now compare simulations with measurements from a radial probe (RRE4) covering the last $9$ turns of the PSI Ring cyclotron. The probe is located $30$ cm
upstream from the $50~\mu$m thick electrostatic extraction septum and, hence gives a very good picture of the beam distribution at the septum.

This probe is able to measure at the full intensity of the $1.3$ MW cw beam. 
In order to compare the simulations with measurements, not only a radial probe is implemented in \opalcycl\ but also all other parameters, described in the previous sections, can be entered into the simulation.

The flattop phase and the injection position and angle are optimized to get the largest turn separation and smallest beam size
at the extraction region, in both the simulation and operation of the PSI Ring cyclotron.

\begin{figure}[H]
   \centering
  \includegraphics*[width=0.95\linewidth]{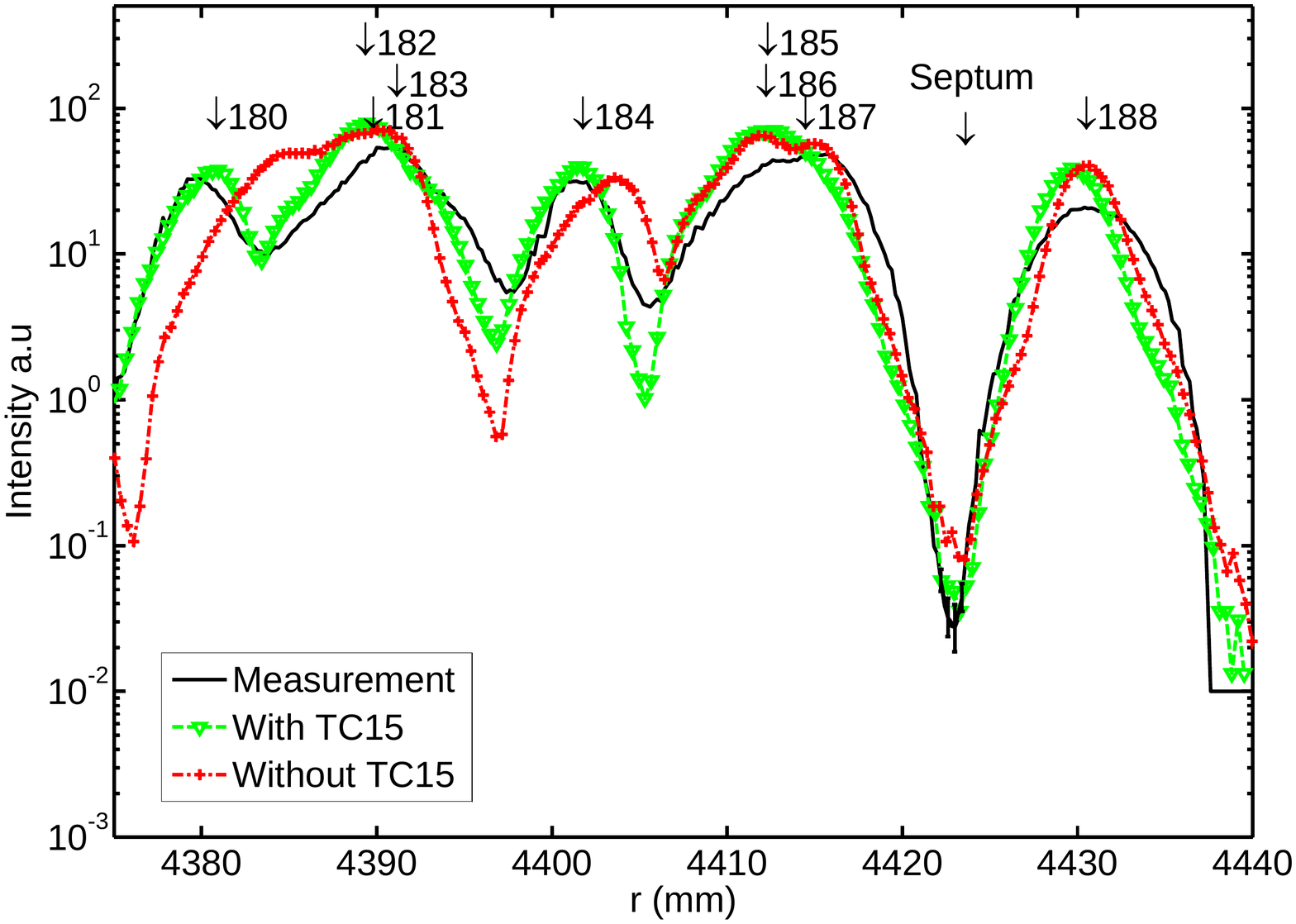}
   \caption{Radial beam profile with indicated turn numbers at extraction for a 2 mA beam, for a parabolic initial distribution at MIC}

   \label{fig:rre4tc15}
\end{figure}

\begin{figure}[H]
   \centering
  \includegraphics*[width=0.95\linewidth]{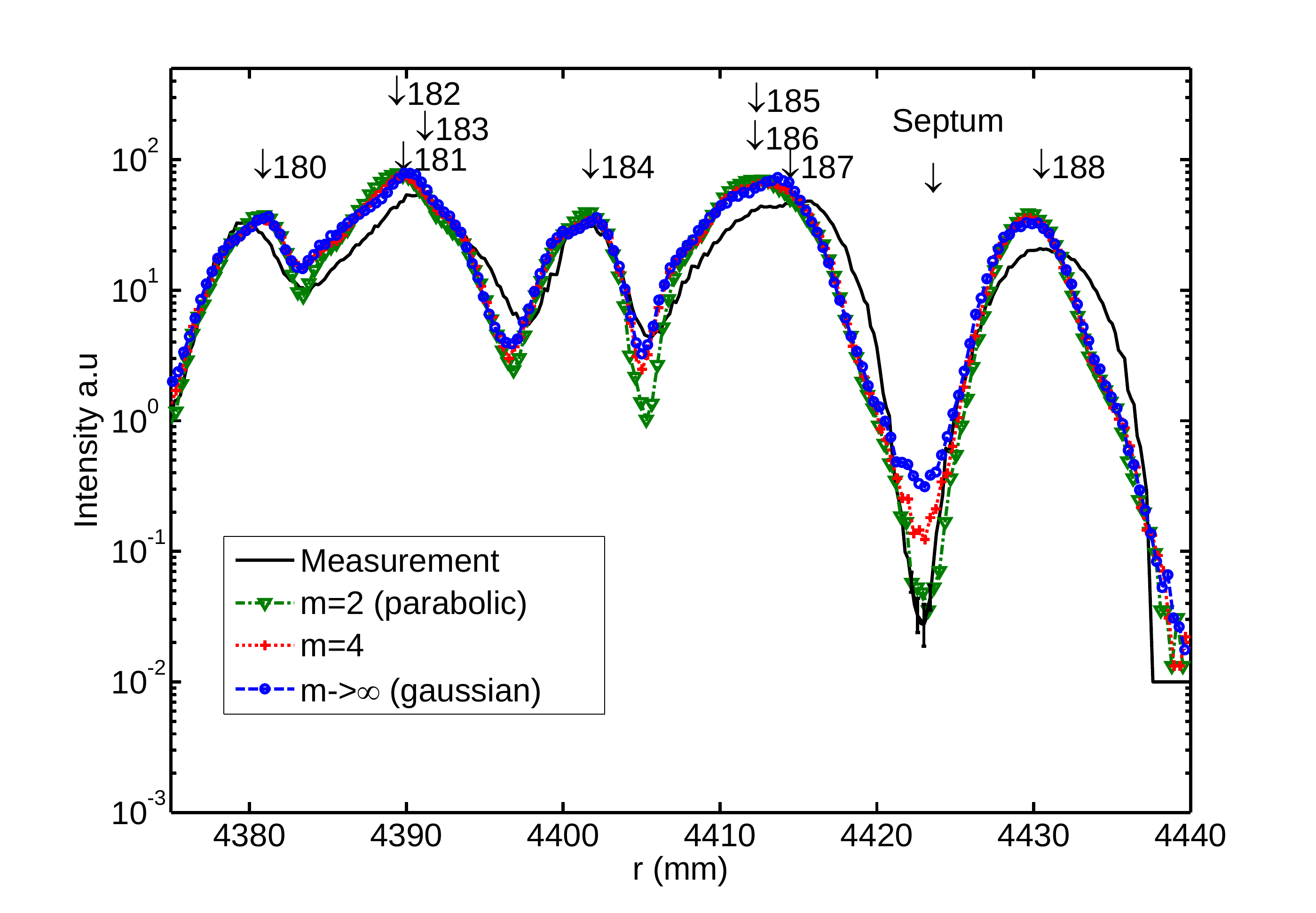}
   \caption{The comparison of different initial beam distributions at the extraction probe, for a 2 mA beam.}
   \label{fig:rre4distr}
\end{figure}

The effect of the trim coil TC15 on the turn pattern is shown in Fig. \ref{fig:rre4tc15}, black denotes the measurement and the colors distinguishing simulations with and without TC15 . For a fixed energy the shift is given by $\frac{\Delta R}{R}=-\frac{\Delta B}{B}$. For turn 180 this shift is:  $\Delta R\mid_{max}\approx 3 mm$, hence the center of turn 180 moves to the exact position of the measurement when considering the effect of the TC15.  
In the PSI cyclotron facility,  the beam is heavily collimated during the early stage of acceleration in the Injector 2 and in the beam transfer line to the PSI Ring cyclotron. As a consequence, the beam
profiles do not follow a Gaussian distribution.
In Fig. \ref{fig:rre4distr} we show again the  intensity pattern at the last 9 turns but for different starting distributions at MIC.
Using the properties of a binomial distribution \cite{TM1114}, we can vary a single parameter $m$ from 0 to $\infty$ and cover a wide range of distributions: from KV to Gaussian.
We find that a parabolic distribution ($m=2$) matches best the measurement at the crucial point, the septum. 
On the extreme side, as expected, the Gaussian distribution (truncated at $3\sigma$) with its tails would fill up the intensity dip and hence would increase the losses at the septum.
This indicated that there are indeed sharp edges in the real distribution, which still differs from the assumed idealized distribution, as suggested by the remaining differences between simulation and measurement (Fig \ref{fig:rre4distr}, e.g at $R=4434$).

Nevertheless, this remarkable agreement is obtained after $\approx 4$km of tracking the millions of macro particles through external- and self fields. This is only possible because of the parallel nature of \opal\
which allows such simulations to be carried out on large high performance computing clusters.  The statistical error of the measurement is indicated at $R=4423$ mm in Fig. \ref{fig:rre4tc15} and \ref{fig:rre4distr}.
At other radii the error bars are significantly smaller and hence not shown in the figures. The statistical errors of the
simulation are smaller than those of the measurement errors due to the large number of simulation particles.


\subsection{SCALING LAW OF BEAM SIZE WITH RESPECT TO  CURRENT}
The energy spread $ \Delta E_{sc}(linear) = e\Delta U_{sc}(linear)$ caused by the linear longitudinal space charge force after $n$ revolutions is given by Joho \cite{Joho} using the sector model:
\begin{eqnarray}\label{eq:dUsc}
  \Delta U_{sc}\approx Z_I \frac{\langle I \rangle}{\Delta \phi/(2\pi)}\frac{n^2}{\beta_f} \text{ with } \\
   Z_I = 2.8 k\Omega = g_{1c}\frac{64\pi}{3} Z_0  \nonumber
\end{eqnarray}
where $g_{1c} \approx 1.4$ (form factor), $Z_0=1/4\pi\eps_0c=30\Omega$ and 
$\langle I \rangle$ is the average current, $\Delta \phi$ is the phase width, $n$ the turn number and $\beta_f=v_f/c$, where $v_f$ is the final velocity of the beam.

The linear energy spread can be compensated with a tilted flattop voltage. This reduces the energy of leading particles and increases the energy of trailing particles.  
There remains a non linear part of the energy spread $\Delta E_{sc}(non linear)$, which can not be compensated.  Let's define 
\begin{equation}
\label{eq:dEnonlin} \Delta E_{sc}(nonlinear) \equiv f_n~ \Delta E_{sc}(linear),
\end{equation}
and note that $f_n$ depends strongly on the beam distribution and is in our case an open parameter in the range of $0.1 \dots 0.5$.

\begin{figure}[H]
   \centering
  \includegraphics*[width=0.90\linewidth]{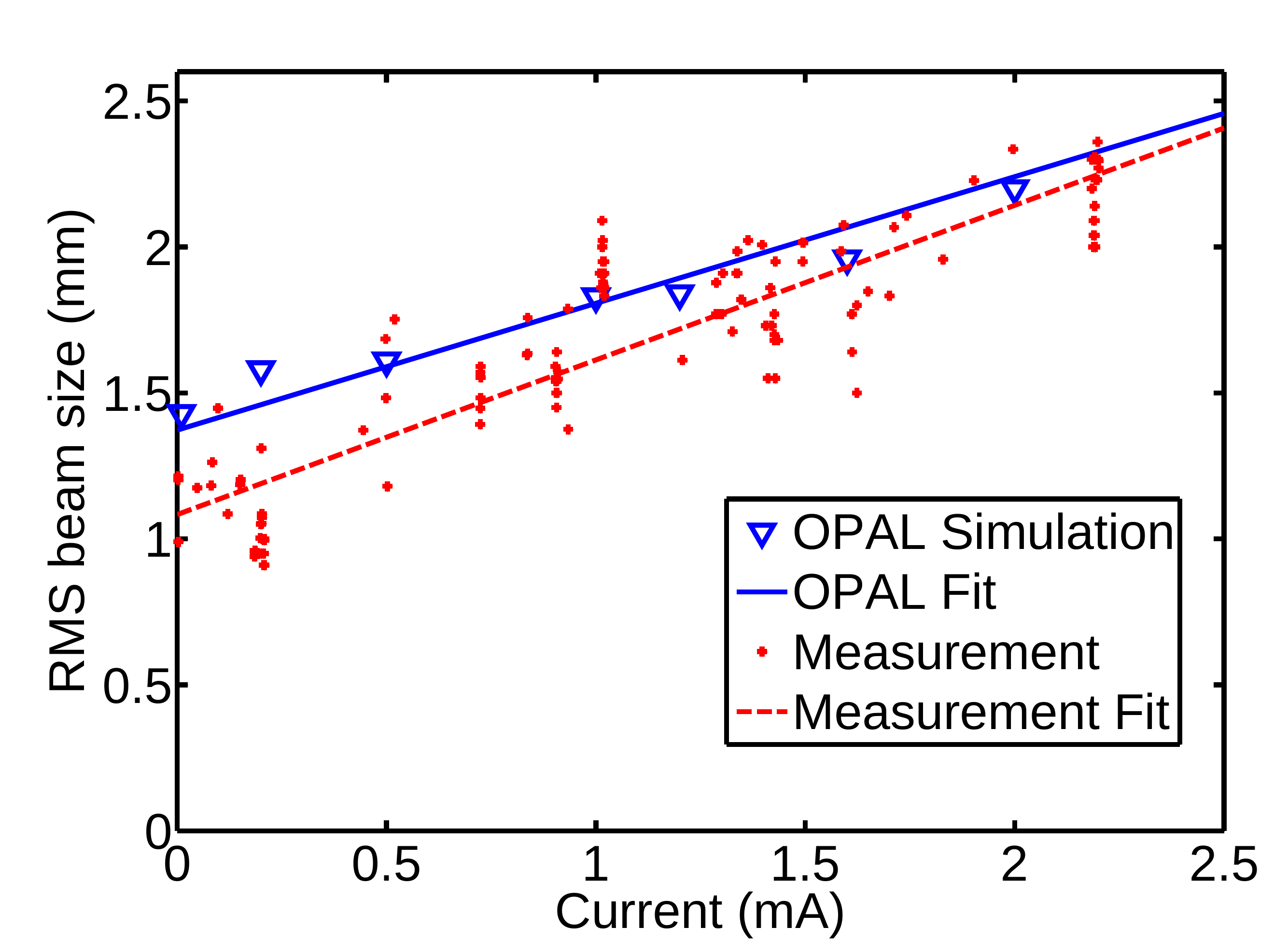}
   \caption{Radial beam size at extraction vs. beam current.}
   \label{fig:rmslongbeam}
\end{figure}

According to Eq. \ref{eq:dRdn}, the space charge induced energy spread leads to a radial spread ($\Delta R_{sc}$) that results in an increase in beam size.
In Fig. \ref{fig:rmslongbeam} we compare beam sizes at the extraction for beam currents from 10 $\mu A$ to 2.2 mA with simulations. 
Even though the measurements where done over a time span of 4 years with very different machine configurations we obtain a good agreement between the
simulations (theory) and the measurements. Hence we can predict very well the extracted beam size as function of the intensity.  This simple model and the precise simulations shown in
the previous paragraph constitutes a benchmarked model for the prediction of the most delicate parameters in  high intensity cyclotrons.

The relation between the average energy gain $dE/dn$ and the turn number $n$ is: 

\begin{equation}
\label{eq:dEdn} n \frac{dE}{dn} = E_f - E_i.
\end{equation}
where $E_f$ is the final energy and $E_i$ is the initial energy.
For single turn extraction the loss on the septum is limited by the ratio 

\begin{eqnarray}
\frac{\Delta E_{sc}(nonlinear)}{\frac{dE}{dn}} \nonumber
\end{eqnarray}
leading to the condition
\begin{equation}
\label{eq:dEineq}\Delta E_{sc}(nonlinear) < \mu_n \frac{dE}{dn}.
\end{equation}

We obtain empirically, for a centered beam a value for  $\mu_n$ which is  $\mu_n \approx 1/3$, where as for an eccentric beam, the turn separation is enhanced, and hence   $\mu_n$ can be as 
high as  $\mu_n \approx 1$.

\begin{figure}[H]
   \centering
  \includegraphics*[width=0.95\linewidth]{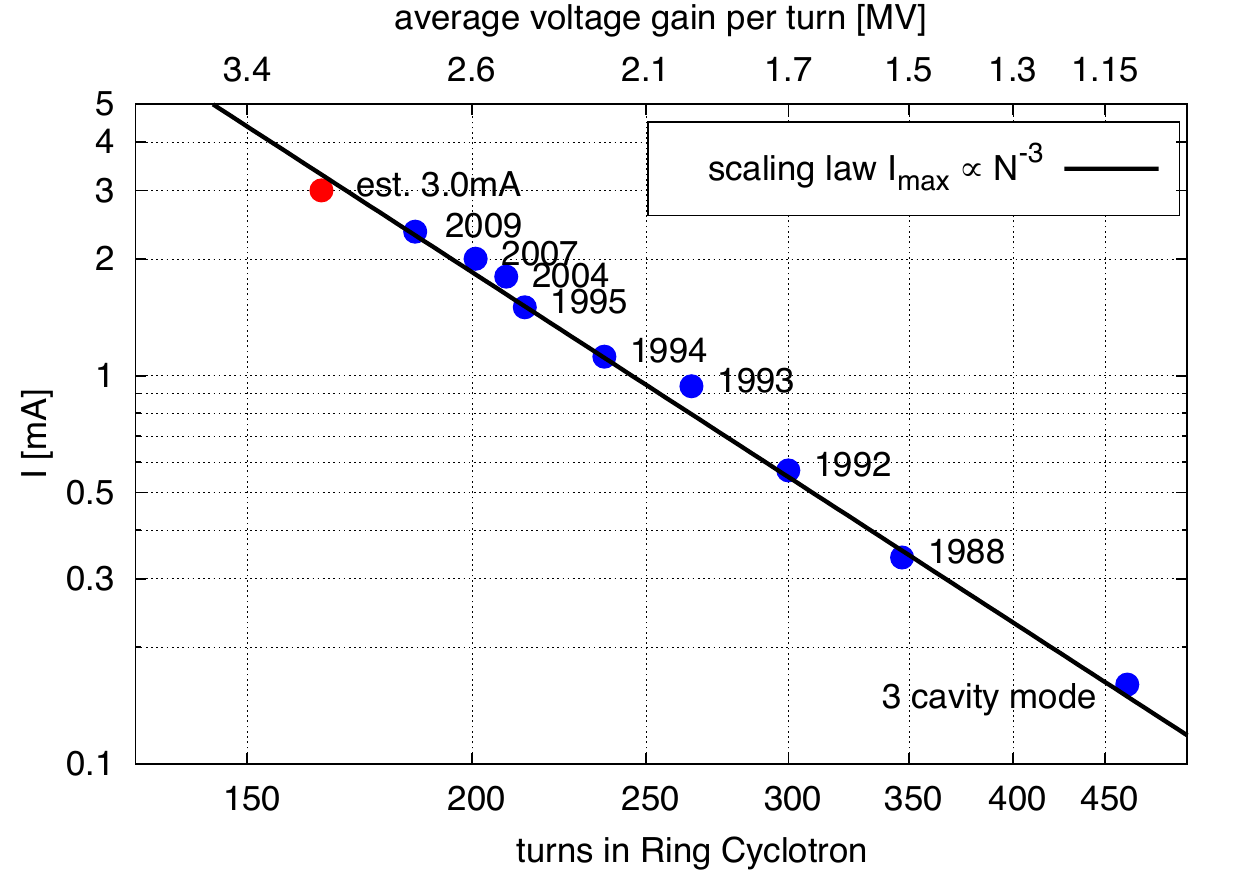}
   \caption{Production beam current versus turn number over a span of 32 years of the PSI Ring cyclotron  } 
   \label{fig:ringcurrent}
\end{figure}

Putting (\ref{eq:dUsc}), (\ref{eq:dEnonlin}), (\ref{eq:dEdn}) and (\ref{eq:dEineq}) together, we get for the current limit from longitudinal space charge forces

\begin{equation}
\label{eq:imax}\ \langle I \rangle_{max} = \frac{\mu_n}{f_n} \frac{U_f - U_i}{Z_I} \frac{\beta_f}{n^3} \frac{\Delta \phi}{2 \pi}.
\end{equation}
where $U_f=E_f/e$ and $U_i=E_i/e$.
Since the turn number $n$ is inverse proportional to the cavity voltage $V_{cav}$, we see the big advantage of a large cavity voltage

\begin{equation}
\langle I \rangle_{max}  \sim  \frac{1}{n^3} \sim V_{cav}^3 
\end{equation}
as experimentally demonstrated by the historical current development in the PSI Ring cyclotron shown in Fig \ref{fig:ringcurrent}.

This relation was first predicted by Joho \cite{Joho} and confirmed during almost 36 years of operation of the PSI Ring cyclotron.

The prediction of the current limit in the PSI Ring cyclotron gives with (\ref{eq:imax}) $\langle I \rangle_{max} = 3 \text{mA}$, using the following parameters:
$U_i=72$ MV, $U_f=590$ MV, $\beta_f=0.8$, $n=188$ turns ($dE/dn = 3$ MeV), $\Delta\phi=12$ and estimating $f_n=1/4$ and $\mu_n=1$ (eccentric injection).

This is remarkably close to the present current limit of $2.3$mA (2010), given the crude assumptions:
\begin{itemize}
\item no turn structure inside the charge sheet (see \cite{Joho} figure 3),
\item non relativistic approximation,
\item no radial boundary condition and 
\item uncertainties in $f_n$ and $\mu_n$.
\end{itemize}
We note that for the PSI Injector 2 (\ref{eq:imax}) is not applicable due to phase mixing \cite{Adam:1}  and hence would give a pessimistic value for the current limit. 



\section{CONCLUSIONS AND DISCUSSIONS}
In this paper, we present  novel precise simulations for the beam dynamics in high intensity cyclotrons. For the first time we are able to obtain a realistic and detailed understanding of the beam dynamics in the very complex PSI Ring cyclotron by means of 3D particle simulations.
By a rough estimation of the initial distribution, according to measurements of beam profile monitors, and the time-structure of the beam, realistic simulations of the PSI Ring cyclotron are presented and compared to measurements. 

Very good agreement for the radial probe between the simulation and measured data is obtained by adjusting the injection position, angle, flattop voltage, and the trim coil TC15. These parameters are all in agreement with settings obtained from the control room.

The presented results with a level of detail large enough to predict limiting tails on the extraction septum (at beam width levels of $3\sigma \dots 4\sigma$), and can be seamless extrapolated to future high power cyclotrons and
enable the precise prediction of crucial parameters, such as losses, based on an existing cw megawatt facility experiences. However a crucial part is the knowledge of the initial distribution from  the evaluation of measurements of beam profiles and time structure.

Primary modeling limitations include an accurate knowledge of the initial particle distribution in the full 6-D phase space, and the lack of particle-matter interaction in our model. Particle-Matter interaction models and resulting struggled primary particles and electrons will 
play an important role when intensity levels increase while at the same time, the losses must be held at present levels. We plan to include these effects in future studies, preliminary results on the particle-matter interaction model are reported in \cite{BiHB2010:1} and ideas for secondary electron creation
and field emission can be found in \cite{WangHB2010:1}.


\section{ACKNOWLEDGMENTS}
The authors thank the Accelerator Modeling and Advanced Simulation (AMAS) group members C.\,Kraus, Y.\,Ineichen and J.\,J.\,Yang for many discussions regarding programming and T.\,Schietinger for providing the post-processing tool
H5PartRoot. We also thank H.\,Zhang for
providing information of the 72 MeV injection line and the PSI Ring cyclotron. This work was partly performed on the {\it felsim} cluster at the Paul Scherrer Institut
and on the Cray XT5 at Swiss National Supercomputing Center (CSCS) within the ``Horizon'' collaboration.

\bibliography{phys-rev-stab-1}

\begin{thebibliography}{16}
\expandafter\ifx\csname natexlab\endcsname\relax\def\natexlab#1{#1}\fi
\expandafter\ifx\csname bibnamefont\endcsname\relax
  \def\bibnamefont#1{#1}\fi
\expandafter\ifx\csname bibfnamefont\endcsname\relax
  \def\bibfnamefont#1{#1}\fi
\expandafter\ifx\csname citenamefont\endcsname\relax
  \def\citenamefont#1{#1}\fi
\expandafter\ifx\csname url\endcsname\relax
  \def\url#1{\texttt{#1}}\fi
\expandafter\ifx\csname urlprefix\endcsname\relax\def\urlprefix{URL }\fi
\providecommand{\bibinfo}[2]{#2}
\providecommand{\eprint}[2][]{\url{#2}}

\bibitem[{\citenamefont{Conrad and Shaevitz}(2010)}]{PhysRevLett.104.141802}
\bibinfo{author}{\bibfnamefont{J.~M.} \bibnamefont{Conrad}} \bibnamefont{and}
  \bibinfo{author}{\bibfnamefont{M.~H.} \bibnamefont{Shaevitz}},
  \bibinfo{journal}{Phys. Rev. Lett.} \textbf{\bibinfo{volume}{104}},
  \bibinfo{pages}{141802} (\bibinfo{year}{2010}).

\bibitem[{\citenamefont{Abderrahim et~al.}(2010)}]{adswhitepaper2010}
\bibinfo{author}{\bibfnamefont{H.~A.} \bibnamefont{Abderrahim}}
  \bibnamefont{et~al.}, \bibinfo{type}{Tech. Rep.}, \bibinfo{institution}{DOE}
  (\bibinfo{year}{2010}), \bibinfo{note}{{Accelerator and Target Technology for
  Accelerator Driven Transmutation and Energy Production}},
  \urlprefix\url{http://www.science.doe.gov/hep/files/pdfs/ADSWhitePaperFinal.%
pdf}.

\bibitem[{\citenamefont{Adelmann et~al.}(2008)\citenamefont{Adelmann, Y.Bi,
  Kraus, Ineichen, Russel, and Yang}}]{opal:1}
\bibinfo{author}{\bibfnamefont{A.}~\bibnamefont{Adelmann}},
  \bibinfo{author}{\bibnamefont{Y.Bi}},
  \bibinfo{author}{\bibfnamefont{C.}~\bibnamefont{Kraus}},
  \bibinfo{author}{\bibfnamefont{Y.}~\bibnamefont{Ineichen}},
  \bibinfo{author}{\bibfnamefont{S.}~\bibnamefont{Russel}}, \bibnamefont{and}
  \bibinfo{author}{\bibfnamefont{J.}~\bibnamefont{Yang}}, \bibinfo{type}{Tech.
  Rep.} \bibinfo{number}{PSI-PR-08-02}, \bibinfo{institution}{Paul Scherrer
  Institut} (\bibinfo{year}{2008}).

\bibitem[{\citenamefont{Adelmann}(2009)}]{ippl:1}
\bibinfo{author}{\bibfnamefont{A.}~\bibnamefont{Adelmann}},
  \bibinfo{type}{Tech. Rep.} \bibinfo{number}{PSI-PR-09-05},
  \bibinfo{institution}{Paul Scherrer Institut} (\bibinfo{year}{2009}).

\bibitem[{\citenamefont{Adelmann et~al.}(2010)\citenamefont{Adelmann, Arbenz,
  and Ineichen}}]{Adelmann20104554}
\bibinfo{author}{\bibfnamefont{A.}~\bibnamefont{Adelmann}},
  \bibinfo{author}{\bibfnamefont{P.}~\bibnamefont{Arbenz}}, \bibnamefont{and}
  \bibinfo{author}{\bibfnamefont{Y.}~\bibnamefont{Ineichen}},
  \bibinfo{journal}{Journal of Computational Physics}
  \textbf{\bibinfo{volume}{229}}, \bibinfo{pages}{4554 }
  (\bibinfo{year}{2010}), ISSN \bibinfo{issn}{0021-9991},
  \urlprefix\url{http://www.sciencedirect.com/science/article/B6WHY-4YHP08T-1/%
2/41309c23eb7fa1b4af95d9401a21da39}.

\bibitem[{\citenamefont{Yang et~al.}(2010)\citenamefont{Yang, Adelmann, Humbel,
  Seidel, and Zhang}}]{PhysRevSTAB.13.064201}
\bibinfo{author}{\bibfnamefont{J.~J.} \bibnamefont{Yang}},
  \bibinfo{author}{\bibfnamefont{A.}~\bibnamefont{Adelmann}},
  \bibinfo{author}{\bibfnamefont{M.}~\bibnamefont{Humbel}},
  \bibinfo{author}{\bibfnamefont{M.}~\bibnamefont{Seidel}}, \bibnamefont{and}
  \bibinfo{author}{\bibfnamefont{T.~J.} \bibnamefont{Zhang}},
  \bibinfo{journal}{Phys. Rev. ST Accel. Beams} \textbf{\bibinfo{volume}{13}},
  \bibinfo{pages}{064201} (\bibinfo{year}{2010}).

\bibitem[{\citenamefont{Baartman}(1995)}]{Baartman:1}
\bibinfo{author}{\bibfnamefont{R.}~\bibnamefont{Baartman}}, in
  \emph{\bibinfo{booktitle}{Proc. 14th Int. Conf. on Cyclotrons and their
  Applications}} (\bibinfo{address}{Capetown}, \bibinfo{year}{1995}), p.
  \bibinfo{pages}{440}.

\bibitem[{\citenamefont{Hockney and Eastwood}(1988)}]{Hockney:1}
\bibinfo{author}{\bibfnamefont{R.~W.} \bibnamefont{Hockney}} \bibnamefont{and}
  \bibinfo{author}{\bibfnamefont{J.~W.} \bibnamefont{Eastwood}},
  \emph{\bibinfo{title}{Computer Simulation Using Particles}}
  (\bibinfo{publisher}{Hilger}, \bibinfo{address}{New York},
  \bibinfo{year}{1988}).

\bibitem[{\citenamefont{Gordon and Taivassalo}(1985)}]{Gordon:2}
\bibinfo{author}{\bibfnamefont{M.~M.} \bibnamefont{Gordon}} \bibnamefont{and}
  \bibinfo{author}{\bibfnamefont{V.}~\bibnamefont{Taivassalo}},
  \bibinfo{journal}{IEEE Trans. Nucl. Sci.} \textbf{\bibinfo{volume}{32}},
  \bibinfo{pages}{2447} (\bibinfo{year}{1985}).

\bibitem[{\citenamefont{D{\"o}lling}(2010)}]{timestructure}
\bibinfo{author}{\bibfnamefont{R.}~\bibnamefont{D{\"o}lling}}, in
  \emph{\bibinfo{booktitle}{Proc. of HB2010}} (\bibinfo{address}{Morschach,
  Switzerland}, \bibinfo{year}{2010}), \bibinfo{note}{"MOPD62"}.

\bibitem[{\citenamefont{Adam and Joho}(1974)}]{TRIMCOIL}
\bibinfo{author}{\bibfnamefont{S.}~\bibnamefont{Adam}} \bibnamefont{and}
  \bibinfo{author}{\bibfnamefont{W.}~\bibnamefont{Joho}}, \bibinfo{type}{Tech.
  Report} \bibinfo{number}{TM-11-13}, \bibinfo{institution}{PSI}
  (\bibinfo{year}{1974}).

\bibitem[{\citenamefont{Joho}(1980)}]{TM1114}
\bibinfo{author}{\bibfnamefont{W.}~\bibnamefont{Joho}}, \bibinfo{type}{Tech.
  Report} \bibinfo{number}{TM-11-14}, \bibinfo{institution}{PSI}
  (\bibinfo{year}{1980}).

\bibitem[{\citenamefont{Joho}(1981)}]{Joho}
\bibinfo{author}{\bibfnamefont{W.}~\bibnamefont{Joho}}, in
  \emph{\bibinfo{booktitle}{9th Int. Conf. on Cyclotrons, p. 337}}
  (\bibinfo{address}{Caen}, \bibinfo{year}{1981}).

\bibitem[{\citenamefont{Adam}(1985)}]{Adam:1}
\bibinfo{author}{\bibfnamefont{S.}~\bibnamefont{Adam}}, \bibinfo{journal}{IEEE
  Trans. on Nuclear Science} \textbf{\bibinfo{volume}{32}},
  \bibinfo{pages}{2507} (\bibinfo{year}{1985}).

\bibitem[{\citenamefont{Bi et~al.}(2010)\citenamefont{Bi, Adelmann, D\"olling,
  Joho, Seidel, Tang, and Zhang}}]{BiHB2010:1}
\bibinfo{author}{\bibfnamefont{Y.~J.} \bibnamefont{Bi}},
  \bibinfo{author}{\bibfnamefont{A.}~\bibnamefont{Adelmann}},
  \bibinfo{author}{\bibfnamefont{R.}~\bibnamefont{D\"olling}},
  \bibinfo{author}{\bibfnamefont{W.}~\bibnamefont{Joho}},
  \bibinfo{author}{\bibfnamefont{M.}~\bibnamefont{Seidel}},
  \bibinfo{author}{\bibfnamefont{C.~X.} \bibnamefont{Tang}}, \bibnamefont{and}
  \bibinfo{author}{\bibfnamefont{T.~J.} \bibnamefont{Zhang}}, in
  \emph{\bibinfo{booktitle}{Proc. of HB2010}} (\bibinfo{address}{Morschach,
  Switzerland}, \bibinfo{year}{2010}), \bibinfo{note}{"TUO2A03"}.

\bibitem[{\citenamefont{Wang et~al.}(2010)\citenamefont{Wang, Adelmann, and
  Ineichen}}]{WangHB2010:1}
\bibinfo{author}{\bibfnamefont{C.}~\bibnamefont{Wang}},
  \bibinfo{author}{\bibfnamefont{A.}~\bibnamefont{Adelmann}}, \bibnamefont{and}
  \bibinfo{author}{\bibfnamefont{Y.}~\bibnamefont{Ineichen}}, in
  \emph{\bibinfo{booktitle}{Proc. of HB2010}} (\bibinfo{address}{Morschach,
  Switzerland}, \bibinfo{year}{2010}), \bibinfo{note}{"MOPD55"}.

\end{thebibliography}

\end{document}